\documentclass[a4paper,twoside]{article}

\usepackage{epsfig}
\usepackage{subfigure}
\usepackage{calc}
\usepackage{amssymb}
\usepackage{amstext}
\usepackage{amsmath}
\usepackage{amsthm}
\usepackage{multicol}
\usepackage{pslatex}
\usepackage{apalike}

\usepackage[utf8x]{inputenc}
\usepackage[T1]{fontenc} 
\usepackage[english]{babel}
\usepackage{hyperref}
\usepackage{float}
\usepackage{caption}
\usepackage{tikz}
\usetikzlibrary{positioning,shapes,arrows,matrix,fit,scopes}
\usepackage{pgfplots}
\usepackage{pgfplotstable}                
\usetikzlibrary{pgfplots.statistics} 
\usepackage{booktabs} 

\usepackage{SCITEPRESS}     

\newcommand{\extn}[1]{{\textsc{#1}}}
\newcommand{\sref}[1]{Section \ref{sec:#1}} 
\newcommand{\fref}[1]{Figure \ref{fig:#1}} 
\newcommand{\eref}[1]{Figure \ref{etext:#1}} 
\newcommand{\lref}[1]{Lemma \ref{lemma:#1}}

\newenvironment{efont}{\fontfamily{\sfdefault}\selectfont}{\par}
\newcommand{\ef}[1]{{\fontfamily{\sfdefault}\selectfont{#1}}}
\newcommand{\ind}{\phantom{x}\hspace{1ex}}
\newcommand{\cf}[1]{{\textcolor{blue}{\bar{#1}}}}

\newcommand{\etext}[3]{
\begin{figure}[H]
\centering
\fbox{%
\begin{minipage}{2.87in}
\begin{efont}
{#3}
\end{efont}
\end{minipage}}
\caption{{#2}}
\label{etext:#1}
\end{figure}
}

\tikzset{
  >= latex,
  el/.style={ellipse, draw, text width=8em, align=center},
  rs/.style={rectangle split, draw, rectangle split parts=#1},
  ou/.style={draw, inner xsep=1em, inner ysep=1ex, fit=#1},
  title/.style={font=\footnotesize, align=center, minimum width=6.7cm},
  proved/.style={line width=.3mm},
  typetag/.style={rectangle, draw=black!50, font=\footnotesize, anchor=west, minimum width=7cm, align=center},
  node distance=.5cm
}

\newtheoremstyle{Elfe}
{.9\baselineskip\@plus.2\baselineskip\@minus.2\baselineskip}
{.9\baselineskip\@plus.2\baselineskip\@minus.2\baselineskip}
{}
{}
{\bfseries}
{.}
{ }
{}

\theoremstyle{Elfe}
\newtheorem{define}{Definition}[]
\newtheorem{lemma}{Lemma}[]

\begin{document}

\title{The \textsc{Elfe} System  \subtitle{Verifying mathematical proofs of undergraduate students} }

\author{\authorname{Maximilian Dor\'e\sup{1} and Krysia Broda\sup{2} 
}
\affiliation{\sup{1}Department of Computing, RWTH Aachen University, Germany
}
\affiliation{\sup{2}Department of Computing, Imperial College London, 180 Queen’s Gate, London SW7 2BZ, U.K.
}
\email{maximilian.dore@rwth-aachen.de, kb@imperial.ac.uk
}
}

\keywords{Didactics of Mathematics, Mathematical Reasoning, Proof Checking, Formal Mathematics}

\abstract{\textsc{Elfe} is an interactive system for teaching basic proof methods in discrete mathematics. The user inputs a mathematical text written in fair English which is converted to a special data-structure of first-order formulas. Certain proof obligations implied by this intermediate representation are checked by automated theorem provers which try to either prove the obligations or find countermodels if an obligation is wrong. The result of the verification process is then returned to the user. \textsc{Elfe} is implemented in \extn{Haskell} and can be accessed via a reactive web interface or from the command line. Background libraries for sets, relations and functions have been developed. It has been tested by students in the beginning of their mathematical studies.}


\onecolumn \maketitle \normalsize \vfill

\section{\uppercase{Introduction}}
\label{sec:introduction}

\noindent The Soviet researcher Victor Glushkov formulated in 1971 that "to understand a proof means to be able to explain it to a machine that is operating with a relatively unsophisticated algorithm" \cite[p. 111]{glushkov}. Remarkably, teaching mathematics in university is still a mostly analogous endeavour. In order to understand mathematical reasoning, students practice writing proofs on paper and wait for the feedback of instructors to improve their understanding. Immediate feedback would greatly increase the learning curve -- it is often difficult to see when a proof is complete or what steps are missing.

Such feedback could be provided by machines. And indeed, many attempts have been made to formalize mathematics. Most prominently, the interactive theorem provers \textsc{Isabelle} and \textsc{Coq} are advanced systems; for instance \textsc{Coq} was used in proving the Four-color-theorem \cite{gonthier}. However, mathematical beginners are overwhelmed by the capabilities of such systems since using them requires a deep understanding of workings of automated theorem provers (ATP).

The goal of this work is to provide users with a system that gives feedback on proofs entered in a fairly natural Mathematical language. Thereby the users are detached from the technicalities of automated theorem provers. The \textsc{Elfe} system provides a proof of concept that this is feasible and sensible. In the past years, several attempts have been made to create a proof verifier which accepts mathematical texts written in fair English, one of which \extn{System for Automated Deduction} (\extn{SAD}) \cite{sad} was most influential for our work. The \extn{SAD} provides an intuitive input language, called \extn{ForTheL}. However, the user still has to dig into the automated verification process to understand why a proof does not work. The \textsc{Elfe} system in contrast processes the output of background provers and tries to give countermodels to wrong proofs.

\etext{example}{Exemplary \textsc{Elfe} text}{
Include functions.\medskip \\
Let A,B,C be set.\medskip \\
Let f: A → B.\\
Let g: B → C.\medskip \\
Lemma: g∘f is injective implies f is injective. \\
Proof: \\
\ind Assume g∘f is injective. \\
\ind Assume x ∈ A and x' ∈ A and (f\{x\}) = (f\{x'\}). \\
\ind Then ((g∘f)\{x\}) = ((g∘f)\{x'\}). \\
\ind Hence x = x'. \\
\ind Hence f is injective. \\
qed.
}

Consider the exemplary proof in \eref{example} which is in fact a valid \textsc{Elfe} text. After including a background library and introducing specific sets \ef{A}, \ef{B} and \ef{C} and functions \ef{f} and \ef{g}, a lemma is proposed that if the composition of \ef{f} and \ef{g} is injective, so the firstly applied \ef{f} must be injective. This lemma is proven by the reasoning that if \ef{f} maps two elements \ef{x} and \ef{x'} to the same element, the composition of \ef{f} and \ef{g} must map them to the same elements. Since this composition is injective, it follows that \ef{x} and \ef{x'} are the same elements and f is thus injective. Note that \ef{ (g∘f)\{x\}} denotes the function application of \ef{g∘f} which is put in brackets to specify the precedence of the symbols. We will learn in the following how the text is verified.

The remainder of the paper is structured as follows. We first give a brief overview of the implementation in \sref{implementation} and web interface in \sref{web}. Next we introduce the Elfe language and proof structures and justify the correctness of the formalisation in \sref{language}. Finally we evaluate our work in \sref{evaluation} and compare it with popular current theorem provers in \sref{relatedWork} before concluding with a short discussion in \sref{futureWork}.

An instance of the system can be found online \footnote{\url{https://elfe-prover.org}}.

\section{\uppercase{Implementation}}
\label{sec:implementation}

\noindent The \textsc{Elfe} system can be accessed through a web interface or a command-line interface (CLI) as shown in \fref{architecture}. The web interface provides an intuitive way of accessing the systems output, while the CLI offers more debugging functionality. We will take a closer look at the the web interface in \sref{web}. 

After the text is entered via one of its interfaces, it will be parsed into an intermediary representation in first-order logic. This proof representation is presented in \sref{statementSequences}. The Verifier takes the intermediary proof representation and checks it for correctness by calling several ATPs in parallel. If a proof obligation is wrong, the Verifier tries to extract a countermodel from the background provers. The result of this verification process is then returned to the user via the chosen interface.

\begin{figure}[h!]
\centering
\tikzstyle{block} = [draw, fill=gray!15, minimum height=3em, minimum width=1.6cm, font=\footnotesize]
\tikzstyle{caption} = [draw,->,font=\footnotesize]
\begin{tikzpicture}[auto, node distance=2cm,>=latex']
    
    \node [block] (command) {CLI};
    \node [block, below=1cm of command] (web) {Web server};

    \node [block, below right=.0cm and .3cm of command] (parser) {Parser};
    \node [block, right=.3cm of parser] (verifier) {Verifier};
    
    \node [block, right=.3cm of verifier] (prover) {ATP};
    
    \draw [caption] (command) -- node {} (parser);    
    \draw [caption] (web) -- node {} (parser);
    \draw [caption] (parser) -- node {} (verifier); 
        
    \draw [caption] (verifier) edge [in=10,out=130] (command);
    \draw [caption] (verifier) edge [in=350,out=230] (web);
    
    \draw [caption] (verifier) -- node {} (prover);
    \draw [caption] (prover) -- node {} (verifier);
    
\end{tikzpicture}
\caption{Architecture of the \textsc{Elfe} system}
\label{fig:architecture}
\end{figure}
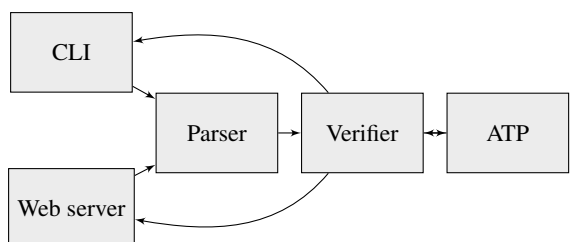

\pagebreak

The system is implemented in \extn{Haskell}, its source code can be found online\footnote{\url{https://github.com/maxdore/elfe}}. In order to parse a text, a parser combinator is constructed with the library \extn{Parsec}\footnote{\url{https://hackage.haskell.org/package/parsec}} \nocite{parsec}. The framework \extn{Scotty}\footnote{\url{https://hackage.haskell.org/package/scotty}} \nocite{scotty} is used to provide a backend for the web interface. The reactive frontend is implemented with the \extn{Javascript} framework \extn{VueJS}\footnote{\url{https://vuejs.org/}} \nocite{vuejs}.
 
In order to send proof obligations to the background provers, the syntax standard \extn{TPTP} \cite{tptp} is used. Since the used ATP can be easily configured, nearly all current systems can be interfaced. 

So far, we have used the provers \extn{E Prover} \cite{eprover}, \extn{SPASS} \cite{spass} and \extn{Vampire} \cite{vampire} due to their performance at the CADE System Competitions \cite{casc}. Additionally, we used the provers \extn{Z3} \cite{z3} and \extn{Beagle} \cite{beagle} which do theorem proving modulo background theories. Even though we did not fully utilize, for instance, their arithmetic proving facilities, it turned out efficient to call several provers in parallel. E.g., \extn{E Prover} turned out to be fast in proving lemmas with equality while \extn{Beagle} gave useful countermodels for wrong proof obligations.

\section{\uppercase{Web interface}}
\label{sec:web}

\noindent The front-end of the web interface shown in \fref{initial} consists of a simple text field in which the user can enter his proof. Above the input, several special characters can be entered by mouse click besides a button that initiates the verification process. 

\begin{figure}[H]
\centering
\includegraphics[width=3in]{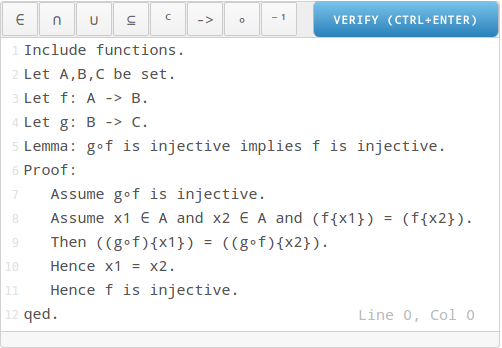}
\caption{The web interface of \textsc{Elfe}}
\label{fig:initial}
\end{figure}

After the verification process has finished, colours indicate the status of each text line as depicted in \fref{verified}. Since all text is green, the text was considered correct. The user can inspect the verification process by clicking in specific lines, more information about the verification is then given in the box below the text field. In our example, we learn the TPTP representation of the proof obligation \ef{x1 = x2} and that it was proved by \extn{E Prover}. Note that the variables are prefixed with \ef{c} in the raw version since they are considered constants at this point in the proof. The reason for this will be explained in \sref{provedStatements}.

\begin{figure}[H]
\centering
\includegraphics[width=3in]{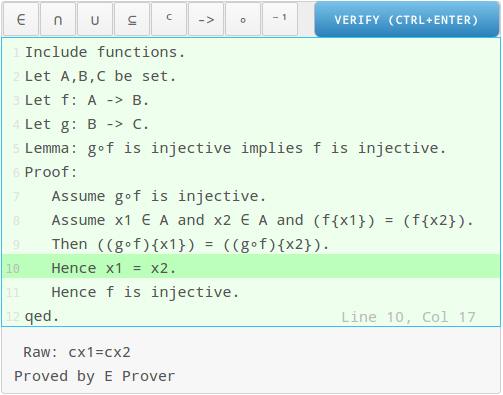}
\caption{Verified correct \textsc{Elfe} text}
\label{fig:verified}
\end{figure}

If the user enters an incorrect proof, as in \fref{wrong}, red colours indicate that the verification process failed. In the example in line 9 we wrongly concluded that \ef{g} must have mapped \ef{x} and \ef{x'} to the same elements, which does not always hold. The background provers could not prove this, but also did not find a countermodel to the obligation. 

\begin{figure}[H]
\centering
\includegraphics[width=3in]{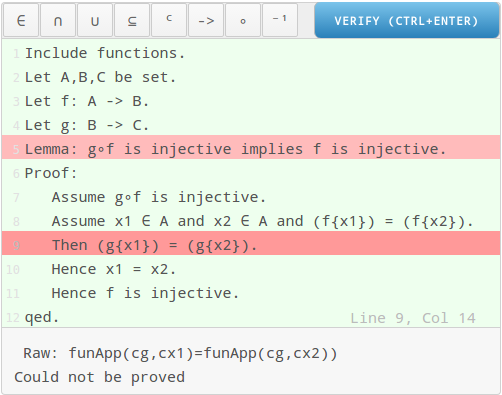}
\caption{An unsound \textsc{Elfe} text}
\label{fig:wrong}
\end{figure}

In the proof in \fref{countermodel}, a countermodel could be found for a wrong conclusion. The lemma states that if a relation \ef{R} is included in \ef{S} and \ef{S} is symmetric, the inverse of \ef{R} must be included in \ef{S} as well. While the statement is in general correct, the proof is too imprecise and misses a case distinction. The countermodel now tells us that if \ef{x} and \ef{y} are in the union of \ef{R} and its inverse, they might be in the inverse of \ef{R} but not in R itself. Thus, the conclusion in line 8 does not in general hold. The correct version of this proof can be found in the Appendix.

\begin{figure}[H]
\centering
\includegraphics[width=3in]{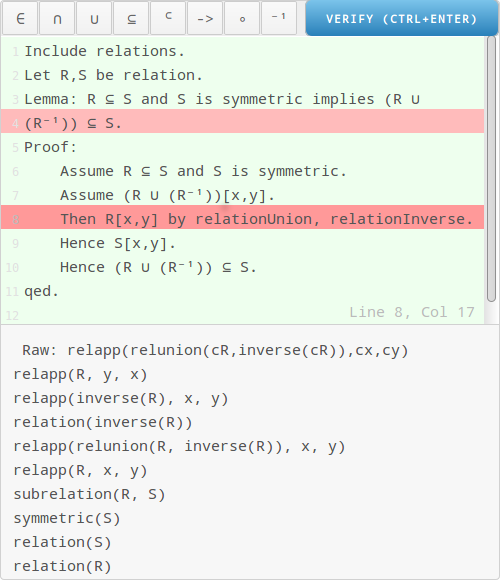}
\caption{Countermodel for a wrong \textsc{Elfe} text}
\label{fig:countermodel}
\end{figure}

\section{\uppercase{\textsc{Elfe} language}}
\label{sec:language}

\noindent The input language for \textsc{Elfe} is mathematical texts written in a subset of natural mathematical language. We will not introduce the whole feature set in this paper and only examine the exemplary proof of \eref{example} in the following. Other language constructs like case distinctions or sub proofs, which make a text less monolithic, are presented in \cite{thesis}.

 In order to verify an \textsc{Elfe} text, we transform it into a special data-structure which implies certain proof obligations. Since this internal proof representation uses first-order logic, we will first introduce how to transform the \textsc{Elfe} language into first-order logic. This preprocessing will be presented in \sref{preproc}. Keywords like \ef{Then} and \ef{Hence} have special meanings in an \textsc{Elfe} proof and are used to structure a mathematical proof. This structure is captured in an intermediate proof representation which is introduced in \sref{statementSequences}. The intermediate proof representation implies certain obligations which need to be checked by the background provers. What these are will be explained in \sref{provedStatements}.

\subsection{From \textsc{Elfe} to First-Order Logic}
\label{sec:preproc}

First-order logic is used to encode mathematical statements. Most transformations are straightforward from \textsc{Elfe} to first-order logic, e.g., \ef{P implies f is injective} is transformed to $P \rightarrow injective(f)$. In order to make an \textsc{Elfe} text more legible, three commands introduce meta-language features.

\etext{functionsLibrary}{Excerpt of the functions library}{
Include sets, relations.\\
Let A,B,C be set. \\
Notation function: f: A → B. \medskip \\
Definition function: for all f.\\
\ind f: A → B iff for all x ∈ A. exists y ∈ B. \\
\ind \ind f[x,y] and \\
\ind \ind (for all y' ∈ B. y = y' or not f[x,y']). \medskip \\
Let f: A → B. \medskip \\
Definition injective: f is injective iff\\
\ind for all x ∈ A, x' ∈ A, y ∈ B. f[x,y] and f[x',y] implies x = x'. \medskip \\
Let g: B → C. \\
Notation composition: g∘f. \\
Definition composition: (g∘f): A → C and \\
\ind (for all x ∈ A. for all y ∈ B. for all z ∈ C. \\
\ind ((f[x,y] and g[y,z]) implies (g∘f)[x,z])).
}

The command \ef{Include} can be used to include the axioms of a background theory. E.g., in our example in \eref{example} we include the functions library with \ef{Include functions}. The user can easily create his own background theory since these are written in the \textsc{Elfe} language as well. You can find an excerpt of the functions library in \eref{functionsLibrary}.

The command \ef{Notation} is used to introduce syntactic sugars. One can write an arbitrary pattern of Unicode characters to define such a pattern, e.g., \ef{Notation function: f: A → B}. The alphabetical parts of the pattern, i.e., \ef{f}, \ef{A} and \ef{B} are treated as placeholders for arbitrary terms. Thus, all terms of the form $$\text{\ef{*: * → *}}$$ with \ef{*} being arbitrary terms are subsequently considered instances of the predicate \ef{function}. For example, \ef{g: B → C} will be transformed internally to the first-order formula $function(g,B,C)$. Similarly, the notation for \ef{composition} is defined as \ef{g∘f}. Consider the version of our proof in raw first-order logic in \eref{exampleRaw}, where the first line of our exemplary \textsc{Elfe} proof \ef{Assume g∘f is injective} is transformed into \ef{Assume} $injective(composition(g,f))$. Note that notations can be used both for term and predicate symbols.

\etext{exampleRaw}{The injectivity proof without syntactic sugar}{
Lemma: $\forall set(A), set(B), set(C), function(f,A,B), $\\$function(g,B,C)$. $injective(composition(g,f)) \rightarrow injective(f)$. \\
Proof: \\
\ind Assume $injective(composition(g,f))$. \\
\ind Assume $funApp(f,x) = funApp(f,x') \\ \ind \ind \ind \ind \ind \ind \wedge in(x,A) \wedge in(x',A)$. \\
\ind Then $funApp(composition(g,f),x) \\ \ind \ind \ind \ind \ind \ind =funApp(composition(g,f),x')$. \\
\ind Hence $x = x'$. \\
\ind Hence $injective(f)$. \\
qed.
}

The command \ef{Let} binds a predicate symbol to a variable, effectively assigning a type to a symbol. By writing \ef{Let A,B, C be set}, we ensure that in all following statements \ef{A}, \ef{B} and \ef{C} have the predicate symbol \ef{set}. Consider \eref{exampleRaw} which shows the injectivity proof after removing meta-level language features. \ef{A}, \ef{B} and \ef{C} are introduced universally quantified as sets in the lemma.

\subsection{Statement Sequences}
\label{sec:statementSequences}

So far, we have only seen how single mathematical statements are transformed into first-order formulas. In order to capture the structure of a proof, we propose a special kind of data-structure, so-called statement sequences. Intuitively, a statement holds a first-order formula with an identifier and a proof. A proof can consist of other statements in order to represent complex proof objects.

\begin{define}\label{def:statement} \textbf{Statement sequences.}\\
A statement $S$ is a tuple $\textsc{Id} \times \textsc{Goal} \times \textsc{Proof}$ where
\begin{itemize}
    \item \textsc{Id} is an alphanumeric string which is unique for each statement
    \item \textsc{Goal} is a formula in first-order logic
    \item \textsc{Proof} is either \\
                \phantom{x}\hspace{3ex} \textsc{Assumed} or  \\
                \phantom{x}\hspace{3ex} \textsc{ByContext} or \\
                \phantom{x}\hspace{3ex} \textsc{BySubContext} \textsc{$Id_1, ... , Id_n$} or \\
                \phantom{x}\hspace{3ex} \textsc{BySequence $S_1, ..., S_n$}  or\\
                \phantom{x}\hspace{3ex} \textsc{BySplit $S_1, ..., S_n$ }
\end{itemize}
A statement sequence is a finite list of statements \textsc{$S_1, ..., S_n$}. 

If a statement $S$ is proved \textsc{BySequence $S_1, ..., S_n$} or \textsc{BySplit $S_1, ..., S_n$}, we call $S_1, ..., S_n$ the children of $S$. If we want to access $S$ from a child $S_i$, we write $S_i$.\textsc{Parent}. On the top level, a statement has no parent, thus $S$.\textsc{Parent} = \textsc{Empty}.
\end{define}

Consider the example in \fref{topLevel}. We will depict a statement visually in the following as a box with its \textsc{Id} in the upper-left corner. The \textsc{Goal} of a statement is written in the header of a statement, the \textsc{Proof} below. A \textsc{Proof} can take different forms to capture complex proof structures. The axioms of a text however are simply annotated by \textsc{Assumed}. E.g., the statements $S_{fun}$ and $S_{inj}$ depict the statements resulting from the definitions in \eref{functionsLibrary}. In the functions library, numerous additional definitions are made which are omitted here. Statements annotated with \textsc{Assumed} will be depicted green in the following. Below the axioms, the statement $S$ of the lemma of our text in \eref{example} follows. In order to prove this statement, we need more advanced proof structures which will be introduced in the next \sref{provedStatements}. The statement is depicted red and with a dashed border to indicate that its proof is not complete.

\begin{figure}[H]
\centering
\begin{center}
\begin{tikzpicture}
\node (S1) [typetag, label={[xshift=-3.2cm, yshift=-0.1cm]\footnotesize $S_{fun}$}, draw=green!80, thick] { $\forall set(A), set(B), f. function(f,A, B) \leftrightarrow \forall x \in A. \exists y \in B. $ \\ $ relapp(f, x, y) \wedge (\forall y' \in B. y = y' \vee \neg relapp(f,x,y')) $ \\ \textsc{Assumed} };

\node (S2) [typetag, below= of S1, label={[xshift=-3.2cm, yshift=-0.1cm]\footnotesize $S_{inj}$}, draw=green!80, thick] { $\forall set(A), set(B), function(f,A,B). $ \\ $ injective(f) \leftrightarrow \forall x \in A, x' \in A, y \in B. $ \\ $ relapp(f,x,y) \wedge relapp(f,x',y) \rightarrow x = x'$ \\ \textsc{Assumed} } edge [<-] (S1);

\node (S) [typetag, below= of S2, label={[xshift=-3.35cm, yshift=-0.1cm]\footnotesize $S$}, draw=red, thick, dashed] {$\forall set(A), set(B), set(C), $ \\ $ function(f,A,B), function(g,B,C)$. \\ $injective(composition(g,f)) \rightarrow injective(f)$ } edge [<-, dotted] (S2);
\end{tikzpicture}
\end{center}
\caption{Exemplary statement sequence}
\label{fig:topLevel}
\end{figure}
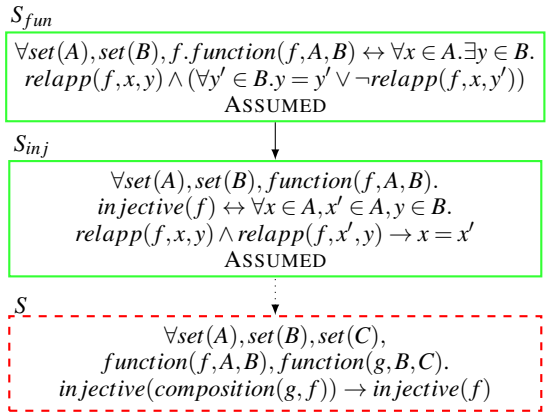

To give an overview of the other types of \textsc{Proof}: A proof \textsc{BySequence} and \textsc{BySplit} makes it possible to nest more complex derivation sequences. A statement annotated with \textsc{ByContext} will be checked by the background provers. \textsc{BySubContext} is a special case of this proof type which allows for restricting the context of the statement.

\subsection{Proved Statements}
\label{sec:provedStatements}

Since we want to verify that a text is sound, we need to introduce a soundness criteria for statements. Axioms of a text are considered correct, but the lemma needs a more subtle criteria.

First we will define which axioms are considered relevant to a statement. Intuitively, the context of a statement in a statement sequence are all statements "above" it. 

\begin{define}\label{def:context} \textbf{Context of a statement}\\
Let $S_1, ... S_n$ be a statement sequence. The context of a statement $S_k$ is inductively defined as
\begin{itemize}
    \item $\Gamma(\textsc{Empty}) = \emptyset$,
    \item $\Gamma(S_k) = \{S_1.\textsc{Goal}, ... , S_{k-1}.\textsc{Goal} \} $ \\ $\qquad \qquad \cup \ \Gamma(S_k.\textsc{Parent})$.
\end{itemize}
\end{define}

For example, in \fref{topLevel}, the context of statement $S$ consists of the respective goals of $S_{fun}$ and $S_{inj}$ (as well as other definitions of the library which are omitted here). With that, we can define an appropriate soundness criteria for statements.

\begin{define}\label{def:proven} \textbf{Proved statement.}\\
Let $S$ be a statement with $S$.\textsc{Goal} = $\phi$. \\We call $S$ proved iff $\Gamma(S) \vDash \phi$. 
\end{define}

In other words, a statement is considered proved if it already followed from the theory created by its context. The statements $S_{fun}$ and $S_{inj}$ in \fref{topLevel} are not proved since they build up the axioms of our theory. The statement $S$ however should follow from these axioms, i.e., should be a proved statement. In order to show that $S$ is proved, we will create a more complex proof object in the following such that correctness of the proof object implies that $S$ followed from its context.

We start by unfolding the outer implication of the lemma $\forall set(A), set(B), set(C), function(f,A,B), $\\$function(g,B,C)$. $injective(composition(g,f)) \rightarrow injective(f)$. More specifically, we fix specific sets \ef{A}, \ef{B} and \ef{C} and functions \ef{f}, \ef{g}. As we see in \fref{assumeLets}, this is captured in our data-structure by introducing another statement $S_1$ such that the proof of $S$ is \textsc{BySequence} $S_1$. We represent proofs \textsc{BySequence} by putting the proof inside the statement to prove. The difference between $S$ and $S_1$ is that we removed the quantifiers and replaced the variables with constants (depicted in blue and with an overline $\cf{A}$ in case the color does not show up).

In order to prove the new goal of $S_1$, we do a so-called unfolding of the implication. The left hand side is put in the statement $S_2$ and annotated with \textsc{Assumed} such that it is in the context of $S_3$, which holds the right hand side of the implication. 

The whole reduction from $S$ to $S_3$ is done automatically by the system. It detects if meta-variables are contained in the goal and injects the proof automatically.

\begin{figure}[H]
\centering
\begin{center}
\begin{tikzpicture}  
\node (S) [title, label={[xshift=-3.5cm, yshift=0.05cm]\footnotesize $S$}] { $\forall set(A), set(B), set(C),  $ \\ $function(f,A,B), function(g,B,C)$.\\ $\qquad injective(composition(g,f)) \rightarrow injective(f)$};

    \node (S1) [below=of S, title, label={[xshift=-3.3cm, yshift=0.05cm]\footnotesize $S_1$}] { $(set(\cf{A}) \wedge set(\cf{B}) \wedge set(\cf{C}) \wedge $ \\ $ function(\cf{f},\cf{A},\cf{B}) \wedge function(\cf{g},\cf{B},\cf{C})) \rightarrow $\\ $ (injective(composition(\cf{g},\cf{f})) \rightarrow injective(\cf{f}))$};

        \node (S2) [below=1.5cm of S1.west, typetag,  label={[xshift=-3.3cm, yshift=-.1cm, align=left]\footnotesize $S_2$}, draw=green!80, thick] {$set(\cf{A}) \wedge set(\cf{B}) \wedge set(\cf{C}) \wedge $ \\ $ function(\cf{f},\cf{A},\cf{B}) \wedge function(\cf{g},\cf{B},\cf{C})$ \\ \textsc{Assumed}};
        
        \node (S3) [below=1.3cm of S2.west, typetag, label={[xshift=-3.3cm, yshift=-.1cm]\footnotesize $S_3$},  draw=red, thick, dashed] {$injective(composition(\cf{g},\cf{f})) \rightarrow injective(\cf{f})$} edge[<-] (S2);

    \node (S1W) [draw=black!50, fit={(S1) (S2) (S3)}] {};

\node [draw=black!50, fit={(S) (S1W)}] {};
\end{tikzpicture}
\end{center}
\caption{Unfolding meta-variables}
\label{fig:assumeLets}
\end{figure}
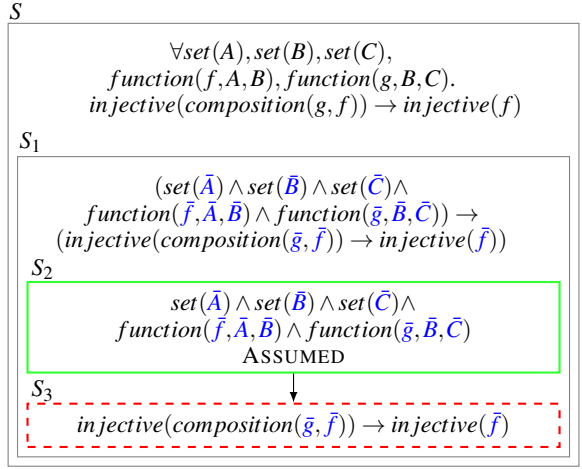

With this, we have reduced the problem of showing that $S$ is proved to showing that $S_3$ is proved. In order to convince us that $S_3$ is proved indeed implies that $S$ is proved, we will first see that it is sound to fix an universally quantified variable to a constant. This can be done by natural deduction which has been shown to be sound \cite{fitting}. Concretely, our construction is analogous to the following deduction rule:

$$ (\forall I): \frac{P(a)}{\forall x. P(x)} \quad \text{with $a$ not occurring in $P(x)$} $$

We use this deduction rule in showing the soundness of our construction in \lref{forallIntroduction}.

\begin{lemma}\label{lemma:forallIntroduction} \ \textbf{$\forall$ introduction.} \\
Let $S$ be a statement such that $S$.\textsc{Goal} = $\forall x. P(x)$ and $a$ not occurring in $S$.\textsc{Goal}, $S$.\textsc{Proof} = \textsc{BySequence} $\boldsymbol{S_1}$, $\boldsymbol{S_1}$.\textsc{Goal} = $P(a)$ and $S_1$ is proved:

\begin{center}
\begin{tikzpicture}  
\node (S) [title, label={[xshift=-3cm, yshift=0.05cm]\footnotesize $S$}, minimum width=6cm] { $\forall x. P(x) \quad (a \text{ not occurring})$ };

    \node (S1) [proved, below=.7cm of S.west, typetag, minimum width=6cm, label={[xshift=-2.9cm, yshift=-.1cm]\footnotesize $\boldsymbol{S_1}$}] {$P(a)$};

\node [draw=black!50, fit={(S) (S1)}] {};
\end{tikzpicture}
\end{center}

\noindent Then $S$ is proved.
\begin{proof}
Since $S_1$ is proved and $\Gamma(S) = \Gamma(S_1)$, we have $\Gamma(S) \vDash P(a)$. With $(\forall I)$ it follows that $\Gamma(S) \vDash \forall x. P(x)$ since $a$ does not occur in $P(x)$.
\end{proof}
\end{lemma}

Next we have to show that it is sound to assume the left hand side of an implication and deduce the right hand side. Again, this is analogous to a natural deduction rule: 

$$ (\rightarrow I): \frac{P \vdash Q}{P \rightarrow Q}$$

This rule is used in the soundness proof in \lref{impliesIntroduction}.

\begin{lemma}\label{lemma:impliesIntroduction} \ \textbf{$\rightarrow$ introduction.}  \\
Let $S$ be a statement such that $S$.\textsc{Goal} = $P \rightarrow Q$, $S$.\textsc{Proof} = \textsc{BySequence} $S_1, \boldsymbol{S_2}$, $S_1$.\textsc{Goal} = $P$, $\boldsymbol{S_2}$.\textsc{Goal} = $Q$ and $S_2$ is proved:

\begin{center}
\begin{tikzpicture}  
\node (S) [title, label={[xshift=-3cm, yshift=0.05cm]\footnotesize $S$}, minimum width=6cm] { $P \rightarrow Q$ };

    \node (S1) [below=.7 of S.west, typetag, minimum width=6cm, label={[xshift=-2.8cm, yshift=-.1cm]\footnotesize $S_1$}] {$P$};
    
    \node (S2) [below=1cm of S1.west, typetag, proved, minimum width=6cm, label={[xshift=-2.8cm, yshift=-.1cm]\footnotesize $\boldsymbol{S_2}$}] {$Q$} edge[<-] (S1);

\node [draw=black!50, fit={(S) (S1) (S2)}] {};
\end{tikzpicture}
\end{center}

\noindent Then $S$ is proved.
\begin{proof}
We have $\Gamma(S_2) = \Gamma(S) \cup \{P\}$. Since $S_2$ is proved, $\Gamma(S) \cup P \vDash Q$. With $(\rightarrow I)$ it follows $\Gamma(S) \vDash P \rightarrow Q$. 
\end{proof}
\end{lemma} 

Now we will construct the proof of $S_3$ as shown in \fref{unfolding}. In the proof text in \eref{example}, we explicitly wrote \ef{Assume $injective(composition(g,f))$. [...] Hence $injective(f)$}. Analogous to the unfolding of the implication of $S_1$ in \fref{assumeLets}, we assume the left hand side and now have to prove the right hand side. Again, this is sound as proved in \lref{impliesIntroduction}.

\begin{figure}[H]
\centering
\begin{center}
\begin{tikzpicture}  
\node (S3) [title, label={[xshift=-3.35cm, yshift=0.05cm]\footnotesize $S_3$}] { $injective(composition(\cf{g},\cf{f})) \rightarrow injective(\cf{f})$};

        \node (S4) [below=1cm of S3.west, typetag, label={[xshift=-3.4cm, yshift=-.1cm]\footnotesize $S_4$}, draw=green!80, thick] {$injective(composition(\cf{g},\cf{f}))$ \\ \textsc{Assumed}};
        
        \node (S5) [below=1.3cm of S4.west, typetag, label={[xshift=-3.4cm, yshift=-.1cm]\footnotesize $S_5$}, draw=red, thick, dashed] {$injective(\cf{f})$} edge[<-] (S4);

\node [draw=black!50, fit={(S3) (S4) (S5)}] {};
\end{tikzpicture}
\end{center}
\caption{Unfolding an implication}
\label{fig:unfolding}
\end{figure}
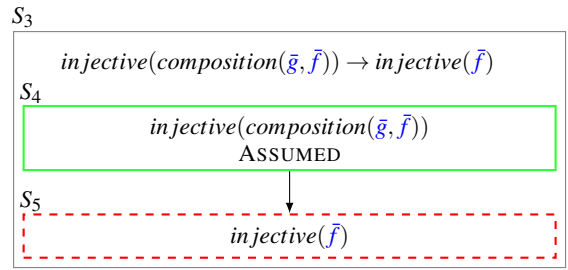

Now, we have to prove that $injective(f)$ holds. In order to do that, the proof in \eref{example} uses the definition of injectivity: \ef{Assume $funApp(f,x) = funApp(f,x') \wedge in(x,A) \wedge in(x',A)$. [...] Hence $x=x'$}. In other words, we prove an alternative goal. In order to retain a sound construction, we have to show two things: First, that the alternative goal indeed implies the original goal and second, that the alternative goal holds. This is represented in \fref{alternativeGoal} by putting two statements $S_6$ and $S_7$ below the goal of $S_5$. This depicts that the \textsc{Proof} of $S_5$ is \textsc{BySplit} $S_6$, $S_7$. Note that a proof \textsc{BySplit} leads to a division of contexts, i.e., he derived goal of $S_6$ will not be put into the context of $S_7$. Thus, the proof \textsc{BySplit} allows for a finer scoping of statements.

The statement $S_6$ contains the soundness check. Its proof is \textsc{ByContext} which means that it will be sent to the background provers. If some ATP finds a proof, a statement annotated with \textsc{ByContext} is considered proved. This is here the case if our definition of injectivity indeed allows us to prove this alternative goal. We will depict statements proved \textsc{ByContext} in orange in the following.

Statement $S_7$ contains the proof of the alternative goal. Again, the universally quantified variables $x$ and $x'$ are fixed to constants. Afterwards, the implication is unfolded. As seen in \lref{forallIntroduction} and \lref{impliesIntroduction}, this is sound.

\begin{figure*}[h!]
\centering
\begin{center}
\begin{tikzpicture}  
\node (S5) [title, minimum width=7.4cm, label={[xshift=-7.1cm, yshift=0.01cm]\footnotesize $S_5$}] { $injective(\cf{f})$};

  \node (S6) [typetag, label={[xshift=-3.3cm, yshift=-.1cm, align=left]\footnotesize $S_6$}, below=.55cm of S5.west, draw=orange!80, thick] {$(\forall x, x'. funApp(\cf{f},x) = funApp(\cf{f},x')  $ \\ $ \wedge in(x,\cf{A}) \wedge in(x',\cf{A}) \rightarrow x = x') \rightarrow injective(\cf{f})$ \\ \textsc{ByContext}};
    
  \node (S7) [title, below right=-1.1cm and .4cm of S6, label={[xshift=-3.4cm, yshift=.01cm]\footnotesize $S_7$}, align=center] { $\forall x, x'. funApp(\cf{f},x) = funApp(\cf{f},x') $ \\ $ \wedge in(x,\cf{A}) \wedge in(x',\cf{A})  \rightarrow x = x'$};
  
    \node (S8) [below=.3cm of S7, title, label={[xshift=-3.3cm, yshift=.04cm]\footnotesize $S_8$}, align=center] { $funApp(\cf{f},\cf{x}) = funApp(\cf{f},\cf{x'}) \wedge in(\cf{x},\cf{A}) \wedge in(\cf{x'},\cf{A}) $ \\ $ \rightarrow \cf{x} = \cf{x'}$ };

        \node (S9) [below=1.1cm of S8.west, typetag, label={[xshift=-3.3cm, yshift=-.1cm, align=left]\footnotesize $S_9$}, draw=green!80, thick] {$funApp(\cf{f},\cf{x}) = funApp(\cf{f},\cf{x'}) \wedge in(\cf{x},\cf{A}) \wedge in(\cf{x'},\cf{A})$ \\ \textsc{Assumed}};
        
        \node (S10) [below=1.2cm of S9.west, typetag, label={[xshift=-3.25cm, yshift=-.1cm]\footnotesize $S_{10}$},  draw=red, thick, dashed] {$\cf{x} = \cf{x'}$} edge[<-] (S9);

    \node (S8W) [draw=black!50, fit={(S8) (S9) (S10)}] {};

  \node (S7W) [draw=black!50, fit={(S7) (S8W)}] {};

\node [draw=black!50, fit={(S5) (S6) (S7W)}] {};
\end{tikzpicture}
\end{center}
\caption{Proving an alternative goal}
\label{fig:alternativeGoal}
\end{figure*}

To convince us that this construction is sound, we have to use two additional natural deduction rules:

$$(\wedge I): \frac{P \quad Q}{P \wedge Q} \qquad (\rightarrow E): \frac{P \rightarrow Q \quad P}{Q}$$

These rules will be used in the proof of \lref{splittingGoal} which is an abstract case of our approach in \fref{alternativeGoal}.

\begin{lemma}\label{lemma:splittingGoal} \ \textbf{Splitting a goal} \\
Let $S$ be a statement such that $S$.\textsc{Goal} = $P$, $S$.\textsc{Proof} = \textsc{BySplit} $\boldsymbol{S_0}, \boldsymbol{S_1}, ... , \boldsymbol{S_n}$, $\boldsymbol{S_0}$.\textsc{Goal} = $Q_1 \wedge ... \wedge Q_n \rightarrow P$, $\boldsymbol{S_i}$.\textsc{Goal} = $Q_i$ and $S_i$ is proved for $i = 1 ,...,n$:

\begin{center}
\begin{tikzpicture}
\node (S) [title, label={[xshift=-3.45cm, yshift=0.05cm]\footnotesize $S$}, minimum width=4.5cm] { $P$ };

    \node (S0) [typetag, minimum width=2.0cm, proved, label={[xshift=-1cm, yshift=-0.1cm]\footnotesize $\boldsymbol{S_0}$}, below=.5cm of S.west] {$Q_1 \wedge ... \wedge Q_n \rightarrow P$};
    \node (S1) [typetag, minimum width=1.9cm, proved, label={[xshift=-.8cm, yshift=-0.1cm]\footnotesize $\boldsymbol{S_1}$}, right=.1cm of S0] {$Q_1$};
    \node (dots) [title, minimum width=.1cm, right=.1cm of S1] {...};
    \node (SN) [typetag, minimum width=1.9cm, proved, label={[xshift=-.8cm, yshift=-0.1cm]\footnotesize $\boldsymbol{S_n}$}, right=.1cm of dots] {$Q_n$};

\node [draw=black!50, fit={(S) (S0) (SN)}] {};
\end{tikzpicture}
\end{center}

\noindent Then $S$ is proved.
\begin{proof}
We have $\Gamma(S) = \Gamma(S_i)$ for $i = 0, \dots , n$. With $S_i$ proved for $i = 1, \dots , n$ we have $\Gamma(S) \vDash Q_i$ for $i = 1, \dots , n$. With $(\wedge I)$ it follows $\Gamma(S) \vDash Q_1 \wedge \dots \wedge Q_n$.

With $S_0$ proved we also have $\Gamma(S) \vDash Q_1 \wedge ... \wedge Q_n \rightarrow P$. Thus, we can deduce with $(\rightarrow E)$ that $\Gamma(S) \vDash P$.
\end{proof}
\end{lemma}

The remaining bit to prove is the goal of $S_{10}$, i.e., that $x = x'$ follows from the context. However, in the text in \eref{example} the next derivation step is \ef{Then $funApp(composition(g,f),x)=funApp(composition(g,f),x')$.}. This statement does not change the overall goal we want to proof, but gives a cornerstone to how one can derive the goal. As depicted in \fref{cornerstone}, this additional finding will first be verified by annotating statement $S_{11}$ with \textsc{ByContext}. Afterwards, the actual goal is proved. Since the user gave no additional proving methods, we send the final goal $x = x'$ to the background provers as well.

\begin{figure}[H]
\centering
\begin{center}
\begin{tikzpicture}  
\node (S10) [title, label={[xshift=-3.25cm, yshift=0.05cm]\footnotesize $S_{10}$}, align=center] { $\cf{x} = \cf{x'}$  };
  
        \node (S11) [below=1cm of S10.west, typetag, label={[xshift=-3.3cm, yshift=-.1cm, align=left]\footnotesize $S_{11}$}, draw=orange!80, thick] {$funApp(composition(\cf{g},\cf{f}),\cf{x})=$ \\ \qquad $funApp(composition(\cf{g},\cf{f}),\cf{x'})$ \\ \textsc{ByContext}};
        
        \node (S12) [below=1.5cm of S11.west, typetag, label={[xshift=-3.3cm, yshift=-.1cm]\footnotesize $S_{12}$}, draw=orange!80, thick] {$\cf{x} = \cf{x'}$ \\ \textsc{ByContext}} edge[<-] (S11);

\node [draw=black!50, fit={(S10) (S11) (S12)}] {};
\end{tikzpicture}
\end{center}
\caption{Giving a cornerstone to a proof}
\label{fig:cornerstone}
\end{figure}

If $S_{11}$ can be derived by the background provers already, the theory created by the context of $S_{12}$ is not extended by adding $S_{11}$. This is formally reflected in \lref{extendingContext}.

\begin{lemma}\label{lemma:extendingContext} \ \textbf{Deriving a cornerstone} \\
Let $S$ be a statement such that $S$.\textsc{Goal} = $P$, $S$.\textsc{Proof} = \textsc{BySequence} $\boldsymbol{S_1}, \boldsymbol{S_2}$, $\boldsymbol{S_1}$.\textsc{Goal} = $Q$, $\boldsymbol{S_2}$.\textsc{Goal} = $P$ and $S_1$ is proved:

\begin{center}
\begin{tikzpicture}
\node (S) [title, label={[xshift=-2.95cm, yshift=0.05cm]\footnotesize $S$}, minimum width=6cm] { $P$ };

   \node (S1) [below=.7cm of S.west, typetag, proved, minimum width=6cm, label={[xshift=-2.85cm, yshift=-.1cm]\footnotesize $\boldsymbol{S_1}$}] {$Q$};
    
    \node (S2) [below=1cm of S1.west, typetag, proved, minimum width=6cm, label={[xshift=-2.85cm, yshift=-.1cm]\footnotesize $\boldsymbol{S_2}$}] {$P$} edge[<-] (S1);

\node [draw=black!50, fit={(S) (S1) (S2)}] {};
\end{tikzpicture}
\end{center}

Then $S$ is proved.

\begin{proof}
Since $S_1$ is proved, we have $\Gamma(S_1) \vDash Q$. Because of $\Gamma(S) = \Gamma(S_1)$ already $\Gamma(S) \vDash Q$. Hence, with $S_2$ proved we have $\Gamma(S_2) \vDash P$ and it follows that already $\Gamma(S) \vDash P$. 
\end{proof}
\end{lemma}

This completes our construction of the internal proof representation of the lemma in \eref{example}. Three statements $S_6$, $S_{11}$ and $S_{12}$ are annotated \textsc{ByContext} and will be sent to the background provers. If each of these three statements can be derived from their respective contexts, we can conclude that the original goal of $S$ already followed from its context. The proof of the lemma is then considered sound.

\section{\uppercase{Evaluation}}
\label{sec:evaluation}

\noindent The tool was tested by students in the beginning of their mathematical studies. In \sref{userFeedback}, we will take a look at their evaluation and suggestions. We also formalized some more advanced theorems in the system, e.g., Cantor's theorem and the Knaster-Tarski theorem, and will discuss our experiences as well as the system's inherent limitations in \sref{limits}.

\subsection{User Feedback}
\label{sec:userFeedback}

\pgfplotstableread{evaluation.dat}{\evaluation}

The system was tested with 12 undergraduates of Computing, Mathematics and Electrical Engineering at Imperial College London, of which none had prior experience with interactive theorem provers. Due to the limited time frame we were not able to evaluate the system further.

At first, the students were given the proof sketch shown in \eref{evalTask1}. An intuition about the proof was given in natural language, i.e., it was explained we want to prove that the complement of the complement of a set is the set itself. All students were able to identify the proof pattern, i.e., that we show set inclusion in both directions. This is a very common proof procedure to show equality of two sets. When writing the remaining bit of the proof, the students successfully resolved syntactic errors by inspecting the parsing errors and all completed the proof. The syntactic characteristics of \textsc{Elfe}, e.g., that \ef{Then} and \ef{Hence} have distinct meanings, did not pose an obstacle since they only had to copy the structure of the first sub proof. However, only two students were able to figure out the meanings of these language features, i.e., that \ef{Hence} closes an implication whereas \ef{Then} is for giving cornerstones to a proof. This suggests that using \textsc{Elfe} requires an introduction to the different language features and users cannot start writing proofs right away.

\etext{evalTask1}{Proof to be completed in the evaluation}{
Include sets.\\
Let A be set.\\
Let x be element.\\
Lemma: ((A\textsuperscript{C})\textsuperscript{C}) = A.\\
Proof:\\
\ind    Proof ((A\textsuperscript{C})\textsuperscript{C}) ⊆ A:\\
\ind\ind        Assume x ∈ ((A\textsuperscript{C})\textsuperscript{C}).\\
\ind\ind        Then not x ∈ (A\textsuperscript{C}).\\
\ind\ind        Hence x ∈ A.\\
\ind    qed.\\
\ind    Proof A ⊆ ((A\textsuperscript{C})\textsuperscript{C}): \\
\ind \ind  ... \\
\ind    qed.\\
qed.
}

Later on, the testers were given more complex proof sketches. Students who were in general comfortable with mathematical reasoning were able to complete the proofs. The other students had problems grasping the idea of the proof and did not start to write a proof in the system.

After the students tried the system, they were given the following statements and had to indicate with 1 (strongly agree) to 6 (strongly disagree) their agreement with the statements.

\begin{itemize}
\item \textit{I enjoy writing mathematical proofs.}\\
      Mean: 3.3 -- Median: 3,5

\item \textit{I find writing mathematical proofs difficult.}\\
      Mean: 2.6 -- Median: 2\\ \\

\item \textit{I think computers can be of use in learning how to write mathematical proofs.}\\
      Mean: 2.3 -- Median: 2 

\item \textit{I enjoyed writing mathematical proofs in the \textsc{Elfe} system.}\\
      Mean: 2.5 -- Median: 2

\item \textit{I found the feedback of the system helpful.} \\
      Mean: 2.6 -- Median: 2

\item \textit{I would like to know how \textsc{Elfe} and interactive theorem proving works.} \\
      Mean: 1.8 -- Median: 1
\end{itemize}

In text form, they could also write down what they liked about the system and what should be improved. It was highlighted that the language was "simple and clear" and did not "get in the way of the proof". They liked the "very understandable and simple UI" and its reactiveness. As improvements for the user interface they proposed autocompletion features of the proofs and syntax highlighting. The given raw translations of the mathematical text were not easy to understand. One user also pointed out that the background provers are sometimes too clever -- thus, a text is accepted even if crucial cornerstones of a proof are missing. He would like to have a criteria on when a proof is "complete" for humans and not only for a computer.

As we see, the testers were in general not especially keen on writing mathematical proofs. Writing proofs in \textsc{Elfe} made it a bit more enjoyable. The system seems to have succeeded in waking an interest for interactive theorem proving.

\subsection{Limits of the Current System}
\label{sec:limits}

Since first-order logic is an intuitive way to write down proofs in set theory and relations, proofs in these domains could be written down easily. Working with the functions library was more complex. Some additional lemmas and function symbols which were introduced to make a proof more readable for humans increase the difficulty for the background provers. If the background provers take too long in proof search, it is hard to assess if a proof itself is wrong or only takes a long time to prove. Debugging a failing proof is still difficult with the user interface provided by \textsc{Elfe}. In most cases, the raw proof obligations given to the background provers were more helpful in finding bugs by manually deleting and changing the given premises. This is due to constructions like \ef{Let} which shorten a proof, but also hide what is going on inside the system.

The \ef{Notation} command has turned out to be a very powerful construct to ease the readability of proofs. New notations can be introduced easily and make a proof look quite intuitive.

\extn{Beagle} was able to provide countermodels to a wrong proof only if the number of premises was limited. Restricting the context of a derivation step increased the success rate significantly. However, for new users it is certainly difficult to relate a countermodel to the entered text since it is given in the raw TPTP format. 

Another problem that occurred was that the background provers were too clever. They sometimes find intermediate steps that are not at all obvious for a human reader. This cleverness is particularly problematic with proofs by contradiction. If the background provers find the inconsistency caused by the assumption, all derivations a user may make are trivially also true, even though they do not make sense in the proof. 

Writing larger proof texts in straightforward domains as set theory can be easily done in
\textsc{Elfe}. However, some properties like well-foundedness are not expressible at all in first-order logic, so it might be expedient for future versions to use higher-order logic at the core of statement sequences.

\section{\uppercase{Related work}}
\label{sec:relatedWork}

\noindent In \sref{verifier}, we will take a look at mathematical text verifiers like the \extn{System for Automated Deduction}, which heavily influenced this project. In \sref{theoremProver}, we will compare \textsc{Elfe} to the popular interactive theorem provers \extn{Isabelle} and \extn{Coq}.

\subsection{Mathematical Text Verifier}
\label{sec:verifier}

In the following, we will present two projects aimed for verifying mathematical texts: The \extn{System for Automated Deduction} (\extn{SAD}) in \sref{sad} and \extn{Naproche} in \sref{naproche}.

\subsubsection{\extn{System for Automated Deduction}}
\label{sec:sad}

The \extn{SAD} was developed at the University Paris and the Taras Shevchenko National University of Kyiv. It continues the project "Algoritm Ochevidnosti" (algorithm of obviousness)  which was initiated by the soviet researcher Victor Glushkov in the 1960s. His goal was to develop a tool that shortens long but "obvious" proofs to users. These omitted parts should be verified by automated theorem provers.  \cite{forthel}

\extn{SAD} uses the input language \extn{ForTheL} which allows for expressing mathematical statements intuitively. \extn{ForTheL} texts are converted to an ordered set of first-order formulas. The structure of the initial text is preserved such that necessary proof tasks can be defined. These tasks are then given to an ATP. The internal reasoner may simplify tasks and omit trivial statements. Afterwards, the verification status of the text is given to the user. For each proof task, the result of the used ATP is returned. This allows to inspect possible sources of failing tasks, but requires knowledge of how the background provers work. \cite{sad} 

Currently, it is not possible to work with functions in \extn{SAD} due to the lack of background libraries. Thus, we could not implement the injectivity proof of \eref{example} in \extn{SAD}.

\subsubsection{ \extn{Naproche}}
\label{sec:naproche}

The \extn{Naproche} system was a joint project between mathematicians at the University of Bonn and linguists at the University of Duisburg-Essen. Its central goal was to develop a controlled natural language (CNL) which checks semi-formal mathematical texts. The input are texts in a \extn{Latex} style language, consisting of mathematical formulas embedded in a controlled natural language.  \cite{naproche}

To extract the semantics of a CNL text, \extn{Naproche} adapts a concept from computational linguistics: Proof Representation Structures (PRS) enrich the linguistic concept of Discourse Representation Structures in such a way that they can represent mathematical statements and their relations. The semantics of PRS have been researched extensively; however, the project is not continued and has no working version available.


\subsection{Interactive Theorem Prover}
\label{sec:theoremProver}

The classical approach to interactive theorem proving integrates a human user strongly in the technical verification process. We will briefly introduce the popular provers \extn{Isabelle} in \sref{isabelle} and \extn{Coq} in \sref{coq} with their respective formalization of the injectivity proof in \eref{example}. \\

\subsubsection{\extn{Isabelle}}
\label{sec:isabelle}

\extn{Isabelle} is a joint project of Cambridge University and the Technical University Munich. It supports polymorphic higher-order logic, augmented with axiomatic type classes. At present it provides useful proof procedures for Constructive Type Theory, various first-order logics, Zermelo-Fraenkel set theory and higher-order logic. \cite{paulson} 

Consider the injectivity proof written in \extn{Isabelle} in \eref{isabelle}. The predicate \ef{inj\_on f A} expresses that function \ef{f} is injective on the domain \ef{A}. The proof structure is close to the one used in \textsc{Elfe}: We introduce arbitrary \ef{x} and \ef{x'} which \ef{f} maps to the same element and conclude that they must have been the same. One has to specify the automated proof tactics and used premises: In our example, the derivations are made by term rewriting using definitions \ef{comp\_def} and \ef{inj\_on\_def} from the background library.

\etext{isabelle}{Proof in \extn{Isabelle}}{
theory InjectiveComposition \\
\ind  imports Fun \\
begin \medskip \\
lemma: \\
\ind  assumes "inj\_on (g ∘ f) A" \\
\ind  shows "inj\_on f A" \\
proof \\
\ind  fix x x' \\
\ind  assume "x ∈ A" and "x' ∈ A" \\
\ind  moreover assume "f x = f x'" \\
\ind  then have "(g ∘ f) x = (g ∘ f) x'" \\ \ind \ind \ind by (auto simp: comp\_def) \\
\ind  ultimately show "x = x'" using assms  \\ \ind \ind \ind by (auto simp: inj\_on\_def) \\
qed
}

In comparison to \textsc{Elfe}, the user is therefore more involved in the automated verification process. Since 2007, \extn{Isabelle} offers the extension \extn{Sledgehammer}. By calling several ATP, \extn{Sledgehammer} tries to determine which premises are important to a goal. It then tries to reconstruct the automated proofs with methods implemented in \extn{Isabelle}. In fact, the mechanical prove methods needed in \eref{isabelle} can be found by invoking \extn{Sledgehammer}.

In a recent study, 34\% of nontrivial goals contained in representative \extn{Isabelle} texts could be proved by \extn{Sledgehammer}. With this extension, \extn{Isabelle} allows beginners to prove challenging theorems. The creators note that \extn{Sledgehammer} was not designed as a tool to teach \extn{Isabelle} since it focused primarily on experienced users. However, it changed the way \extn{Isabelle} is taught. Beginners do not have to learn about low level proving tactics and how they work but can focus on the proof from a higher level. \cite{threeYears}

\subsubsection{\extn{Coq}}
\label{sec:coq}

\extn{Coq} is an interactive theorem prover initially developed 1984 at INRIA. It is based on the Curry–Howard correspondence which relates types to classical logic. In order to prove a proposition, one has to construct a term with the type corresponding to the proposition.

Consider the injectivity proof implemented in \eref{coq}. Again, the idea of the proof is to show that \ef{f x = f x'} implies \ef{x = x'}. However, we have to explicitly apply rewrite techniques to make the derivation steps. The tactic \ef{intuition} says that we can assume a left hand side of an implication and then prove the right hand side. Afterwards, we want to make sure that we can just apply \ef{g} on both sides. We have to rewrite both sides of \ef{H0}, which stands for \ef{f x = f x'}, in order to get to our assertion. The final goal \ef{x = x'} is then derived by applying the rewrite technique \ef{auto}.

\etext{coq}{Proof in \extn{Coq}}{
Require Import Basics. \\
Definition injective \{A B\} (f : A → B) := \\
\ind  forall x y : A, f x = f y → x = y. \\
Theorem c\_inj (A B C:Type) (f:A→B) (g:B→C): \\
\ind  (injective (compose g f)) → injective f. \\
Proof. \\
\ind    intuition. \\
\ind    intros x x'. \\
\ind    pose (f x = f x'). \\
\ind    intuition. \\
\ind    assert (g (f x) = g (f x')). \\
\ind    \{  elim H0.  rewrite H0.  trivial. \}     \\
\ind auto. \\
Qed.
}

As we see, the translation process of mathematical texts to functional programs requires a good understanding of type theory and is not suitable for mathematical beginners.

Consequently, the most prominent current interactive theorem provers are of a deeply technical natural. They are thought of as programming languages that happen to prove theorems, and not digitisations of mathematical language.

\section{\uppercase{Discussion}}
\label{sec:futureWork}

\noindent This paper presented \extn{Elfe}, a system that checks proofs in discrete mathematics. Entered texts are transformed to statement sequences, a special data-structure of first-order formulas. Remaining proof obligations are then checked by background provers. Statement sequences are a powerful intermediate proof representation which can hold manifold proof techniques. The clear soundness criteria allows for extending the proof techniques easily.

Students who tested the system liked especially that they got immediate feedback on their proof work. The implemented background libraries allow for an easy start. Once a user becomes familiar with the tool, he can easily construct his own background libraries. Certainly, more evaluation of the system in pedagogical environments is necessary. It will be particularly interesting to examine how teachers can incorporate the system in their courses.

The language constructs presented here were the result of formalizing several exemplary proofs. If one formalizes more proofs, he will probably feel the need for additional proving methods. If one can map the proving methods soundly into statement sequences, this should be easy to implement. 

In addition to giving countermodels for wrong proofs, one could utilize more features of the background provers. Many provers return in-depth information about the proof of a conjecture. This information could be useful for users in order to understand why a proof works or fails. The challenge is to present the technical output of the background provers via an intuitive interface. In order to do proofs with arithmetic, it might be useful to utilize already implemented arithmetic capabilities of background provers such as \extn{Z3} and \extn{Beagle}. Expert users presumably prefer systems with deep insight into the technical verification process, but an abstraction is necessary if we want to use computers in teaching mathematics.

The biggest structural limitation of \textsc{Elfe} is that it internally uses first-order logic. E.g., with the current capabilities it is not straightforward to implement proofs by induction. The recent years have seen interesting advances in automated theorem proving of typed higher-order logic. A new standard for typed higher-order-logic has been added to \extn{TPTP} which is used by several provers like \extn{Leo-II} \cite{leo} and \extn{Satallax} \cite{satallax}. A next version of \textsc{Elfe} could use this development in order to provide a more powerful way of expressing mathematics. This requires to introduce a meaningful type system for \textsc{Elfe}.

\vfill
\bibliographystyle{apalike}
{\small
\bibliography{paper}}

\section*{\uppercase{Appendix}}
\label{appendix}

\etext{coq}{Correct \textsc{Elfe} proof about relations}{
Include relations. \\
Let R,S be relation. \\
Lemma: R ⊆ S and S is symmetric implies\\ \ind (R ∪ (R⁻¹)) ⊆ S. \\
Proof:\\
\ind    Assume R ⊆ S and S is symmetric.\\
\ind    Assume (R ∪ (R⁻¹))[x,y].\\
\ind    Then R[x,y] or (R⁻¹)[x,y].\\
 \ind   Case R[x,y]:\\
 \ind\ind       Then S[x,y] by subrelation.\\
 \ind   qed.\\
  \ind  Case (R⁻¹)[x,y]:\\
   \ind\ind     Then R[y,x] by relationInverse.\\
   \ind\ind     Then S[y,x] by subrelation.\\
    \ind\ind    Then S[x,y] by symmetry.\\
   \ind qed.\\
   \ind Hence S[x,y].\\
   \ind Hence (R ∪ (R⁻¹)) ⊆ S.\\
qed.
}

\vfill
\end{document}